\documentclass[aps,prl,amsmath,amssymb,superscriptaddress,showkeys,showpacs,twocolumn,floatfix]{revtex4-2}

\usepackage{amsmath,amssymb,mathtools}	
\usepackage{graphicx,color,subfigure}
\usepackage{float}
\usepackage{appendix}
\usepackage{hyperref}
\usepackage{booktabs}

\begin{document}

\title{Chiral magnetic effect amplified baryogenesis at first-order phase transitions}

\author{Hui Liu}
\affiliation{Department of Physics and Chongqing Key Laboratory for Strongly Coupled Physics, Chongqing University, Chongqing 401331, P. R. China}

\author{ Ligong Bian }\email{lgbycl@cqu.edu.cn}

\affiliation{Department of Physics and Chongqing Key Laboratory for Strongly Coupled Physics, Chongqing University, Chongqing 401331, P. R. China}

\begin{abstract}
In this study, we show that, in the background of the primordial magnetic field, the chiral magnetic effect can significantly amplify the chiral chemical potential sourced by the CP violation near the bubble walls during the first-order electroweak phase transition. This effect can lift the generated baryon asymmetry by several orders, and make it possible to explain the baryon asymmetry of the Universe with a CPV in the fermion sector far beyond the limitation of the electron dipole moments.

\end{abstract}

\maketitle

\section{Introduction} 
Among many baryogenesis mechanisms, the electroweak baryogenesis (EWBG) is well motivated to explain the generation of baryon asymmetry of the Universe (BAU), since its two crucial ingredients' detectability~\cite{vandeVis:2025efm, Morrissey:2012db}, i.e., the new physics driving the first-order phase transition can be probed at colliders~\cite{FCC:2018byv, Arkani-Hamed:2015vfh} and gravitational wave detectors such as Laser Interferometer Space Antenna (LISA)~\cite{LISA:2017pwj, Caldwell:2022qsj, Athron:2023xlk}, and colliders and electron dipole moment (EDM) searches can detect the required CP-violations (CPV) beyond the standard model. Firstly, to generate enough magnitude of net baryon number requires a slow bubble velocity with a thick wall. In contrast, a high magnitude gravitational wave signature produced from the first-order electroweak phase transition usually requires fast-moving bubble walls after nucleation~\cite{Bian:2021ini}. Secondly, to generate a large chiral asymmetry requires a large CP-violation source, which excludes many new physics models from electron dipole momentum measurements, such as ACME~\cite{ACME:2018yjb}, except in cases where different CP-violation sources cancel each other out~\cite{Bian:2014zka}. 

In this situation, the lepton-mediated EWBG takes advantage of the top in the sense that: the CP violations in the Higgs-lepton Yukawa couplings are less constrained by LHC and EDM experiments~\cite{Alonso-Gonzalez:2021jsa, vandeVis:2025efm, Fuchs:2020uoc, Xie:2020wzn}. However, 
since the CPV couplings involving $\tau$ lepton can contribute to the two-loop Barr-Zee diagram~\cite{Brod:2013cka},
the recent electron EDM constraint from the ACME reads $d_e/e\leq 1.1\times 10^{-29} $e cm at 90\% C.L.~\cite{ACME:2018yjb}, which limits the NP scale of the CPV operator: $\Lambda_f\gtrsim 0.16$ TeV. Ref.~\cite{Fuchs:2020uoc} shows that, considering the limit of eEDM and that of LHC, successful interpretation of the BAU with electroweak baryogenesis requires: $\Lambda_f\lesssim 0.87$ TeV. 
Under the MFV, recent studies further show that~\cite{Alonso-Gonzalez:2021jsa}:  $s_\tau v^2/\Lambda_f^2\lesssim 1.7 \times 10^{-3}$ considering eEDM limits, which can set $\Lambda_f\gtrsim 5.97 $ TeV for $s_\tau\sim\mathcal{O}(1)$, and therefore shut the window to explain the BAU with the traditional EWBG. 

Primordial magnetic fields generated during some physical processes in the early Universe are thought to be the seed of the cosmological magnetic field that exists in the voids of large-scale structures~\cite{Dolag_2010, Tavecchio_2010, Tavecchio_2011, Vovk_2012}, the Milky Way~\cite{Wielebinski2005}, and galaxy clusters~\cite{Clarke_2001, Bonafede_2010, Feretti_2012}, and whose strength at finite correlation length can be probed
by gamma-ray observations of blazars at present~\cite{Dermer_2011, Taylor_2011, Neronov_2010} after considering its evolution in the early Universe ~\cite{Grasso_2001,Widrow_2002,Kulsrud_2008,Kandus_2011, Widrow_2011,Durrer_2013}. 
Early studies show that the evolution of a primordial (hyper)magnetic field with helicity would be largely affected by 
a preexisting lepton asymmetry through the chiral magnetic effect (CME)~\cite{Vilenkin:1980fu, Kharzeev:2004ey, Kharzeev:2007tn, Kharzeev:2007jp, Joyce:1997uy, Campbell:1992jd, Boyarsky:2011uy, Boyarsky:2012ex, RostamZadeh:2015xnd, Boyarsky:2015faa, Semikoz:2003qt, Semikoz:2004rr, Semikoz:2005ks, Semikoz:2009ye, Akhmetev:2010qdw, Dvornikov:2011ey, Semikoz:2012ka, Dvornikov:2012rk, Semikoz:2013xkc, Semikoz:2015wsa, Semikoz:2016lqv, Pavlovic:2016mxq}, and the evolution of the modified magnetohydrodynamics (MHD) would further amplify the lepton asymmetry through the chiral anomaly~\cite{Joyce:1997uy, Giovannini:1997eg, tHooft:1976rip}. The resulting lepton asymmetry can be further transferred to the baryon asymmetry through the electroweak sphaleron process and further explain the 
observed BAU before the electroweak phase transition~\cite{Fujita:2016igl, Abbaslu:2021mkt, Semikoz:2013xkc, Domcke:2022uue, Semikoz:2016lqv, Kamada:2016eeb, RostamZadeh:2015xnd, Anber:2015yca}, and Ref.~\cite{Kamada:2016cnb} presents a more systematic study by including the electroweak crossover effect. 
In this work, we consider the local $\tau-$CPV around the vacuum bubbles during the first-order phase transition to generate the chiral potential of lepton, and find that such a lepton asymmetry can be significantly enhanced by the CME effect around the bubbles. This effect can increase the generated baryon asymmetry by around  $\mathcal{O}(10^2-10^3)$.

\section{The mechanism}

%------------------------------------------------------------

We model the Higgs background across the wall by a standard kink profile~\cite{DeVries:2018aul}
\begin{equation}
  \phi_b(z)
  = \frac{v}{2}\Bigl(1 + \tanh\frac{z}{L_w}\Bigr),
  \qquad
  \label{eq:phib_profile}
\end{equation}
where $v$ is the zero-temperature Higgs vev and $L_w$ denotes the wall
thickness.
During the electroweak phase transition in the presence of a bubble, the effective fermion mass is spacetime-dependent. We split $\bar{m}_{\tau}=m_{\tau}(\phi_b,T)+m_{f,T}(T)$, with $m_{f,T}(T)$ the usual finite-temperature mass and~\cite{DeVries:2018aul}
\begin{equation}
  m_\tau(z)
  = \frac{\phi_b(z)}{\sqrt{2}}
    \left[
      y_\tau
      + i\,\frac{\phi_b^2(z)}{2\Lambda_f^2}
    \right]\;,
  \label{eq:mtau_complex}
\end{equation}
where $y_\tau$ is the SM Yukawa coupling and $\Lambda_f$ denotes the cutoff scale of
the dimension-six CPV operator with  $\mathcal{L}_{CPV}=-\frac{m_\tau}{v}h\bar{\tau}\tau-\frac{s_\tau m_\tau}{\Lambda_f^2}v h\bar{\tau}i \gamma_5\tau$, which would be constrain by the EDM
experiments. 
The CPV source entering the transport equation can be written as
\begin{equation}
  S_{\rm CP}(z)
  = s_f\,v_w\,N_c\,
    \frac{1}{\pi^2}\,
    \left[
      \frac{y_\tau^2}{2\Lambda_f^2}\,
      \phi_b^3(z)\,\phi_b'(z)
    \right] J_\tau\;,
  \label{eq:SCP_general}
\end{equation}
where $v_w$ is the wall velocity, $N_c$ counts the number of colors of the
fermion species (here $N_c=1$ for $\tau$), $s_f=\pm1$ is chosen to obtain a net number of baryons (rather than antibaryons), $J_{\tau}$ is a numerical factor whose expression is presented in Eq.~\eqref{Jtau} of Appendix A.

With the CP-violating source term, we have a space-dependent chemical potential for the $\tau-$lepton. We define the chiral chemical potential $\mu_5$ via the axial charge density $n_5=n_{\tau_R}-n_{\tau_L}$ and relate $n_5$ to $\mu_5$:  $n_5/\mu_5=k_{\tau}T^2/6$ with $k_{\tau}=1$.
In the background of the primordial magnetic field, the transportation equation for the chemical potential of the lepton takes the form of:
\begin{equation}
  \frac{\partial\mu_5}{\partial t}
  - v_w\,\frac{\partial\mu_5}{\partial z}
  - D_{\rm eff}\,\frac{\partial^2\mu_5}{\partial z^2}
  = \alpha\,\boldsymbol{E}\!\cdot\!\boldsymbol{B}
    - \tilde{\Gamma}_{\rm tot}\,\mu_5
    + \tilde{S}_{\rm CP},
  \label{eq:mu5-final}
\end{equation}
where we have defined
\begin{equation}
  D_{\rm eff} \equiv \frac{6D_{\tau}}{T^2},\qquad
  \alpha \equiv \frac{6}{T^2}\,\frac{g'^2}{16\pi^2},\qquad 
  \end{equation}
\begin{equation}
  \tilde{\Gamma}_{\rm tot} \equiv \frac{6}{T^2}\,\Gamma_{\rm tot},\qquad
  \tilde{S}_{\rm CP} \equiv \frac{6}{T^2}\,S_{\rm CP}\;.
\end{equation}
Here, $D_{\tau}=100/T_n$ is the diffuse constant for right-handed $\tau$, $\Gamma_{\mathrm{tot}}=(\Gamma
_M+\Gamma_Y)(\frac{1}{k_{\tau}}+\frac{1}{k_l})$, $\Gamma_M$ and $\Gamma_Y$ denote the helicity-flipping and Yukawa rates, respectively; their detailed definitions as well as the  temperature-dependent coefficients $k_i$ are given in Appendix~A~\cite{Xie:2020wzn}. We focus on the plasma in the symmetric phase (in front of the bubble wall). Therefore, throughout this work $\mathbf{E}\equiv\mathbf{E}_Y$
and $\mathbf{B}\equiv\mathbf{B}_Y$ denote hypermagnetic fields.
 Eq.~\eqref{eq:mu5-final} is the desired evolution equation in the planar-wall approximation, including diffusion,
advection, damping, the anomalous $\boldsymbol{E}\!\cdot\!\boldsymbol{B}$
source from the background of the primordial hypermagnetic field. 

When the chiral chemical potential appears, one has an additional CME-induced contribution to the MHD equation of the magnetic field, 
\begin{equation}
  \frac{\partial\boldsymbol{B}}{\partial t}
  = \frac{1}{\sigma}\,\boldsymbol{\nabla}^2\boldsymbol{B}
    + \boldsymbol{\nabla}\times(\boldsymbol{v}\times\boldsymbol{B})
    + \frac{\alpha_Y}{\pi\sigma}\,
      \boldsymbol{\nabla}\times(\mu_5 \boldsymbol{B}) \;,
  \label{eq:B-vector-eq08}
\end{equation}
where $\bf v$ is bulk velocity, $\alpha_Y=g'^2/4\pi\simeq0.01$. The first term on the right-hand side describes the resistive diffusion of the
magnetic field. For the electroweak theory, the conductivity of plasma is $\sigma=70~{\rm T}$~\cite{Joyce:1994zn}. The second term encodes plasma advection and stretching of
field lines, and the third term is the CME-induced contribution, which can
Lead to an instability and exponential growth of helical magnetic fields.

We now rewrite Eq.~\eqref{eq:B-vector-eq08} in component form for the same
one-dimensional planar geometry in the wall frame. For the detailed derivation of these results, we refer the reader to Appendix B.
\begin{align}
  \frac{\partial B_x}{\partial t}
  &= \frac{1}{\sigma}\,\frac{\partial^2 B_x}{\partial z^2}
     + v_w\,\frac{\partial B_x}{\partial z}
     - \frac{\alpha_Y}{\pi\sigma}\,
       \frac{\partial}{\partial z}(\mu_5 B_y),
  \label{eq:Bxfinal9}\\[0.5em]
  \frac{\partial B_y}{\partial t}
  &= \frac{1}{\sigma}\,\frac{\partial^2 B_y}{\partial z^2}
     + v_w\,\frac{\partial B_y}{\partial z}
     + \frac{\alpha_Y}{\pi\sigma}\,
       \frac{\partial}{\partial z}(\mu_5 B_x).
  \label{eq:Byfinal10}
\end{align}
and $\partial_z B_z=0$. Here, we considered the transverse velocities of the magnetic field plasma vanish $v_x=v_y=0$ and $v_z=-v_w$ since we work in the wall frame.
Eqs.~\eqref{eq:mu5-final}, \eqref{eq:Bxfinal9}, and \eqref{eq:Byfinal10}
constitute the closed system for the coupled evolution of the chiral chemical
potential and the transverse magnetic field in the planar-wall geometry,
which is used in our numerical analysis.

The CME effect would significant grow the PMF around the bubble wall during the first-order electroweak phase transition through Eqs.~(\ref{eq:Bxfinal9},\ref{eq:Byfinal10}), and further amplify the chiral chemical potential through Eq.~\eqref{eq:mu5-final}, which further yield enhencement of the baryon number generated in front of the wall. With the chemical potential of the baryon number governed by 
\begin{align}
\frac{\partial \mu_B}{\partial t}  - v_w\,\frac{\partial \mu_B}{\partial z} 
&= D_B\,\frac{\partial^2\mu_B}{\partial z^2}
-\frac{15}{4}\Gamma_{ws}\,\mu_B
+ C_B\,\Gamma_{ws}\,\mu_5,
\label{eq:muB-final11}
\end{align}
Here, $D_B=\frac{6D_q}{T^2}$,  $D_q=6/T$ is the quark diffusion constant,  $C_B =219/89$, $\Gamma_{ws}=6\kappa \alpha_w^5T$ is the EW sphaleron rate with $\kappa\sim 20$ and $\alpha_w=g^2/(4\pi)$~\cite{Bodeker:1999gx, Moore:1999fs, Moore:2000mx}. One can calculate the baryon asymmetry $\eta_B\equiv n_B/s= T^2\mu_B/6s$ where $s=2\pi^2g_{\star}T^3/45$ is the entropy density and Standard Model degree of freedom $g_{\star}=106.75$.

For a tau-lepton CPV source using a semiclassical approach, 
without taking into account CME and the magnetic field, the final baryon asymmetry within the traditional EWBG can be obtained as~\cite{DeVries:2018aul}: 
\begin{equation}
\begin{aligned}
    Y_B=&\frac{3\Gamma_{ws}}{2v_ws}\frac{2\bar{S}}{\sqrt{4D_l\bar{\Gamma}_B+v_w^2}+\sqrt{4D_l\bar{\Gamma}_S+v_w^2}}\times\\
    &\frac{L_{ws}L_S}{L_S+\frac{1}{2}L_{ws}(1+\sqrt{1+\frac{4D_q}{L_{ws}v_w}})}\;,
\label{YB1811}
\end{aligned}
\end{equation}
where $D_l=100/T$ is left-handed $\tau$ lepton diffuse constant,
$L_i=\frac{2D_l}{(v_w+\sqrt{4D_l\bar{\Gamma}_i+v_w^2})}$, $\bar{S}=\int_0^{\infty}\exp(-y/L_B)S_{\tau}dy$, $L_{ws}=v_w/(\mathcal{R}\Gamma_{ws})$ with $\mathcal{R}=15/4$ being the SM relaxation
term, and $i={S,B}$ indicated as symmetric and broken phase respectively. 

\section{Numerical results}

\begin{figure}[!htp]
\centering
\includegraphics[width=0.9\linewidth]{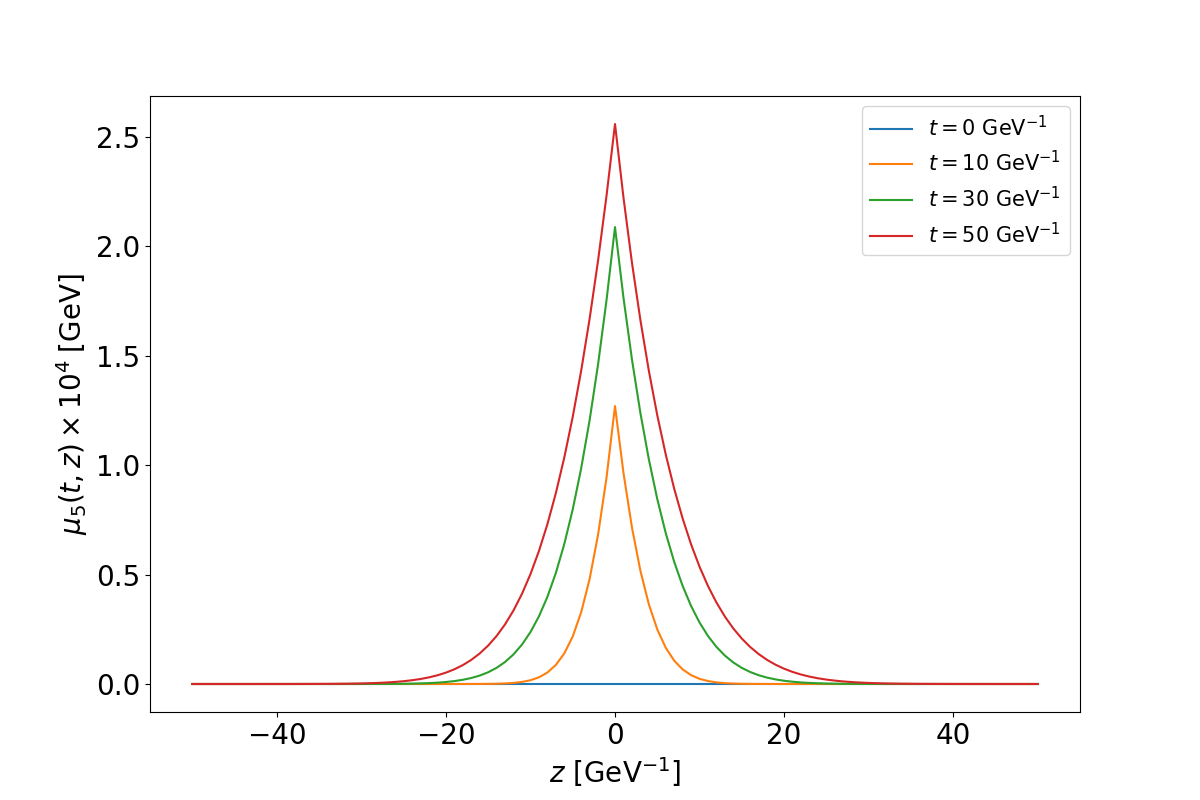}
\includegraphics[width=0.78\linewidth]{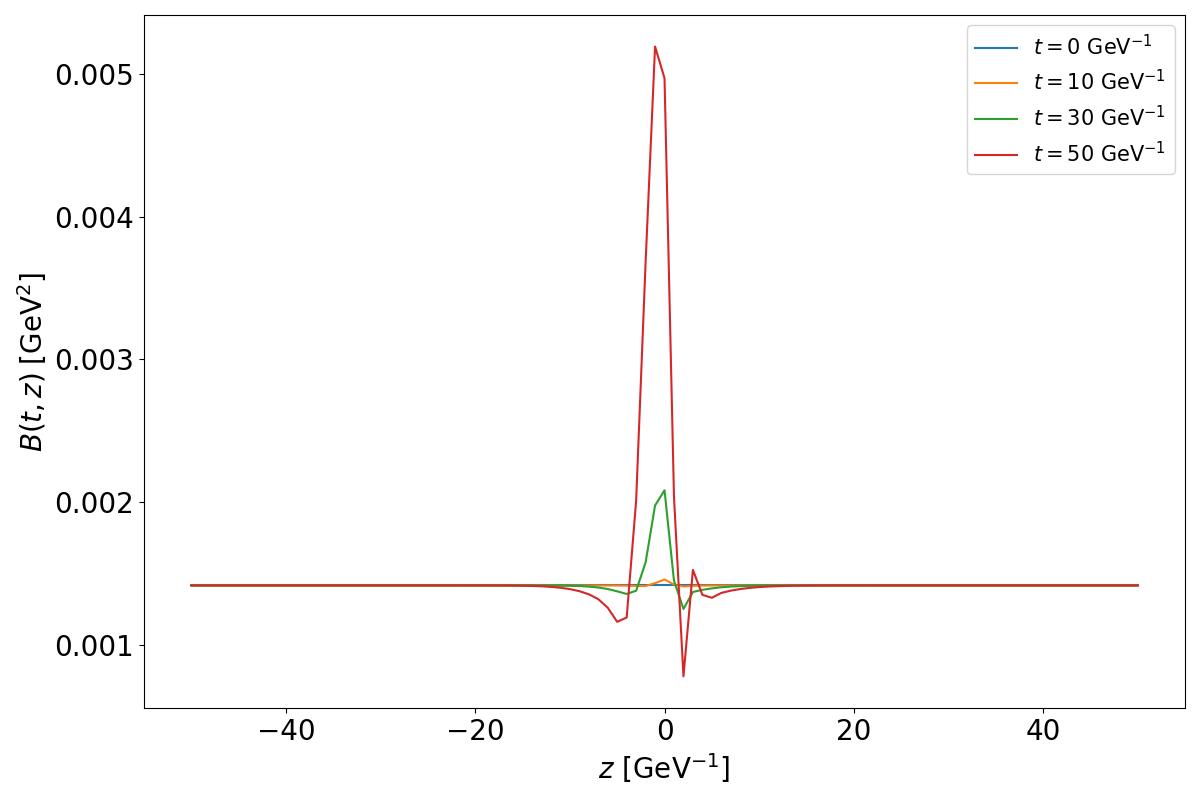}
\caption{Time evolution of the chiral chemical potential and magnetic field near the electroweak bubble wall. Upper: Spatial profiles of the chiral chemical potential $\mu_5(t,z)$, generated by the CPV $\tau$-sector source localized on the bubble wall. Lower: Corresponding evolution of the transverse local hypermagnetic field magnitude ${B}(t,z)=\sqrt{B_x^2(t,z)+B_y^2(t,z)}$ including the CME near the wall in the symmetry phase. Results are shown for several representative times $t=0,10,30,50\;\mathrm{GeV}^{-1}$ with the wall centered at $z=0$.}
\label{mu5}
\end{figure}

To obtain the properties of the chemical potential and magnetic field,  we numerically solve the coupled transport-MHD equations, i.e., Eqs.~(\ref{eq:mu5-final},\ref{eq:Bxfinal9},\ref{eq:Byfinal10}), on $z\in[z_{min}, z_{max}]$ and impose $\mu_5(t, z_{min/max})=0$, $\partial_zB_{x,y}(t,z_{min/max})=0$, and the axial chemical potential $\mu_5(t,z)$ is initially set to zero $\mu_5(0,z)=0$.  the primordial magnetic field is initially considered as homogenesis with $B_x(0,z)=B_y(0,z)=10^{-3}\;\mathrm{GeV}^2$. This corresponds to $B_p(T_n)\simeq 5\times10^{16}\mathrm{G}$ and hence an equivalent present-day amplitude $B_0^{today}\simeq 7\times10^{-18}\mathrm{G}$ after considering the relation between the electroweak epoch physical strength and the present-day amplitude as $B_p\simeq10^{20}\mathrm{G}(T/100\mathrm{GeV})^{7/3}(B_0^{today}/10^{-14}\mathrm{G})\times \mathcal{G}_B(T)$, where $B_0^{today}$ is present-day magnetic field, and $\mathcal{G}_B$ is $\mathcal{O}(1)$ factor that depend on the number of relativistic species~\cite{Kamada:2016eeb}, here we considered the inverse-cascade scaling for a maximally helical hypermagnetic field. 
This value is comfortably below CMB upper limits on primordial magnetic fields (typically nG at 1 Mpc), so our choice of $B_Y^0$ is conservative. We stress that our simulation evolves a local near-wall hypermagnetic configuration; the $(B_0^{today}, \lambda
_0)$ mapping is used only as an order of magnitude consistency check. The time variable $t$ in our transport-MHD equations denotes the local microphysical evolution time in the plasma near the bubble wall. Cosmological expansion can be neglected because the integration interval satisfies $t_{\mathrm{max}}\ll H(T_n)^{-1}$ (equivalently $t_{\mathrm{max}}\ll t_{\mathrm{cos}}(T_n)\simeq M_0/2T_n^2 $), where $t_{\mathrm{cos}}$ is cosmological time, and $M_0=\sqrt{90/(8\pi^3g_{\star})}M_{pl}$, $g_{\star}=106.75$ is the effective number of relativistic degrees of freedom, $M_{pl}$ is Planck mass. Full numerical details are provided in Appendix B.

Figure~\ref{mu5} illustrates the real-time generation and evolution of chiral asymmetry and magnetic fields in the vicinity of a planar electroweak bubble wall moving with velocity $v_w=0.05$ at nucleation temperature $T_n=88\;{\mathrm{GeV}}$, with the wall centered at $z=0$: $L_w=0.11 \mathrm{GeV}^{-1}$ in Fig.~\ref{mu5}, see also Ref.~\cite{Semikoz:2009ye, Liu:2024mdo,Di:2025ncl,Di:2024gsl} to generate BAU with the magnetic fields around this magnitude. 
The upper plot shows that,  at $t=0$, the axial charge density vanishes everywhere, and the CPV source quickly generates a localized chiral asymmetry near the wall. Owing to the dependence $S_{\mathrm{CP}}\propto \phi_b^3\phi_b'$,  the source term is sharply localized around the wall center at $z=0$, leading to the rapid formation of a localized chiral asymmetry. And, as shown in the bottom plot, the transverse magnetic field, initialized as a homogeneous background $B_Y^0=\sqrt{B_x^2+B_y^2}$, develops a localized distortion around the wall once $\mu_5$ is generated. This behavior is driven by the CME term in the induction equation, $\nabla \times (\mu_5 {\bf B})$, which couples the spatial gradient of $\mu_5$ to the transverse field components. As time evolves, the anomaly term proportional to ${\bf E}\cdot {\bf B}$ further enhances the magnitude of the $\mu_5$. As the $\mu_5$ grows with time, the CME contribution acts as an effective ``$\alpha$-term" that amplifies the magnetic field in the vicinity of the wall, producing a peak whose structure reflects the derivative coupling $\partial_z(\mu_5 B_{x,y})$ in Eq.~\eqref{eq:Bxfinal9}, \eqref{eq:Byfinal10}.
Since the diffusion and damping are not yet efficient enough to spread the asymmetry into the bulk, so $\mu_5$ rapidly approaches a quasi-steady, $\mu_5$ approaches a quasi-steady profile determined by the balance between the CPV source, diffusion, advection with the wall velocity $v_w$, the total flipping rate $\Gamma_{\mathrm tot}$, and the anomaly term. 
 Far away from the wall, where $\mu_5\rightarrow 0$, the magnetic field remains close to its initial value, while diffusion smooths the profile and limits the growth at late times. The chiral asymmetry remains concentrated near the wall, reflecting the short diffusion length relative to the wall width. The sharply peaked lepton asymmetry sources baryon number via the weak sphaleron transitions behind the wall.

\begin{figure}[!htp]
\centering
\includegraphics[width=0.8\linewidth]{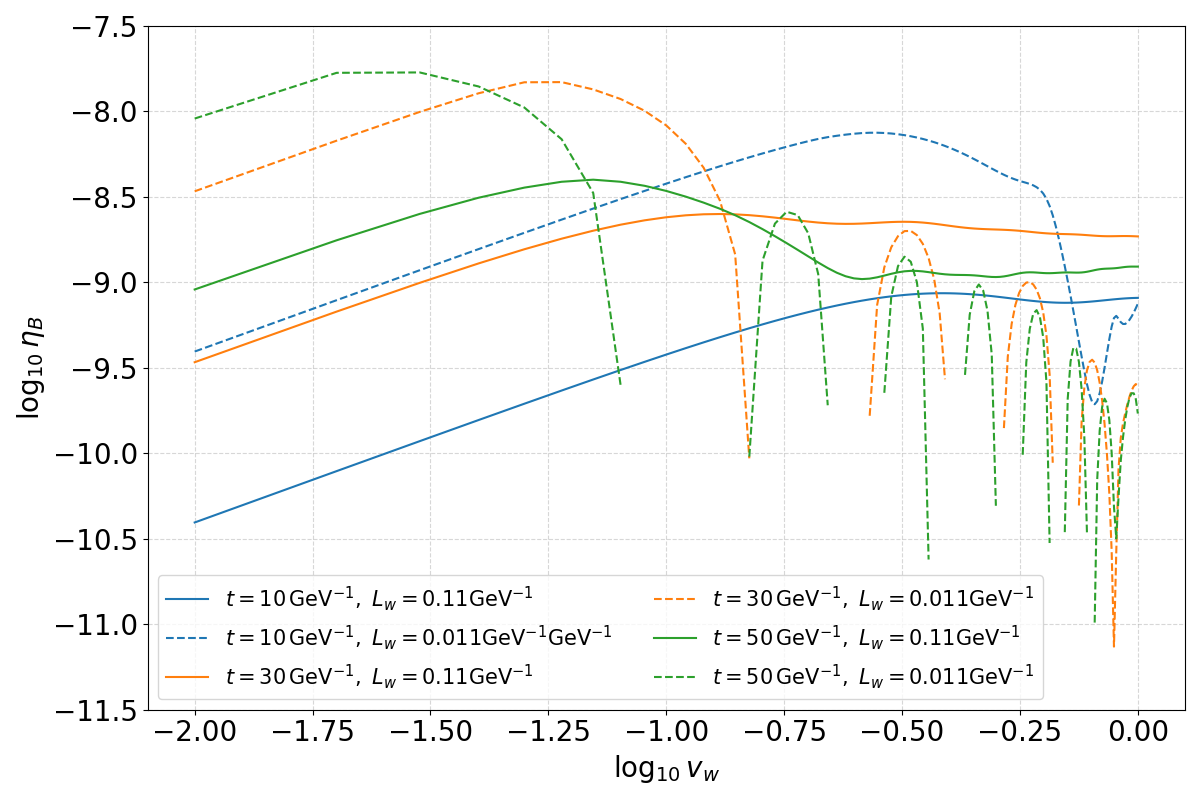}
\caption{Bubble wall velocity dependence of the baryon asymmetry in CME amplified electroweak baryogenesis, the y/x-axis is plotted on a base-10 logarithmic scale.  The three color-matched pairs of curves correspond to snapshots at $t=10, 30, 50,\; \mathrm{GeV^{-1}}$. Solid (dashed) curves use a thick (thin) wall profile with $L_w=0.11\;\mathrm{GeV^{-1}} (L_w=0.011\; \mathrm{GeV^{-1}})$. The CPV source is generated in the $\tau$-sector with a cutoff $\Lambda_f=1000\;\mathrm{GeV}$, and the baryon number is sourced through the weak sphaleron response to $\mu_5$ encoded in the $\mu_B$ equation.}
\label{eta_vw}
\end{figure}
Figure~\ref{eta_vw} shows the dependence of the BAU ($\eta_B=n_B/s$) on the bubble wall velocity $v_w$, highlighting the interplay between transport dynamics, CPV sources, and CME-induced magnetic effects, after solving Eq.~\ref{eq:muB-final11} together with Eqs.~(\ref{eq:mu5-final},\ref{eq:Bxfinal9},\ref{eq:Byfinal10}). The nontrivial $v_w$ dependence reflects competing physical effects. For small wall velocities, the CPV source remains active over a longer effective time, allowing chiral asymmetry to diffuse and be efficiently converted into baryon number by weak sphalerons. As $v_w$ increases, advection dominates and the effective interaction time decreases, suppressing the baryon yield. At the same time, the CME-induced amplification of magnetic fields enhances the chiral asymmetry at later times, partially compensating for this suppression. The comparison between different wall thicknesses demonstrates that thinner walls lead to more localized and stronger CP-violating sources, modifying both the magnitude and the velocity dependence of $\eta_B$. Overall, Figure~\ref{eta_vw} shows that the inclusion of CME effects can significantly alter the standard velocity dependence expected in conventional electroweak baryogenesis.

\begin{figure}[!htp]
\centering
\includegraphics[width=0.8\linewidth]{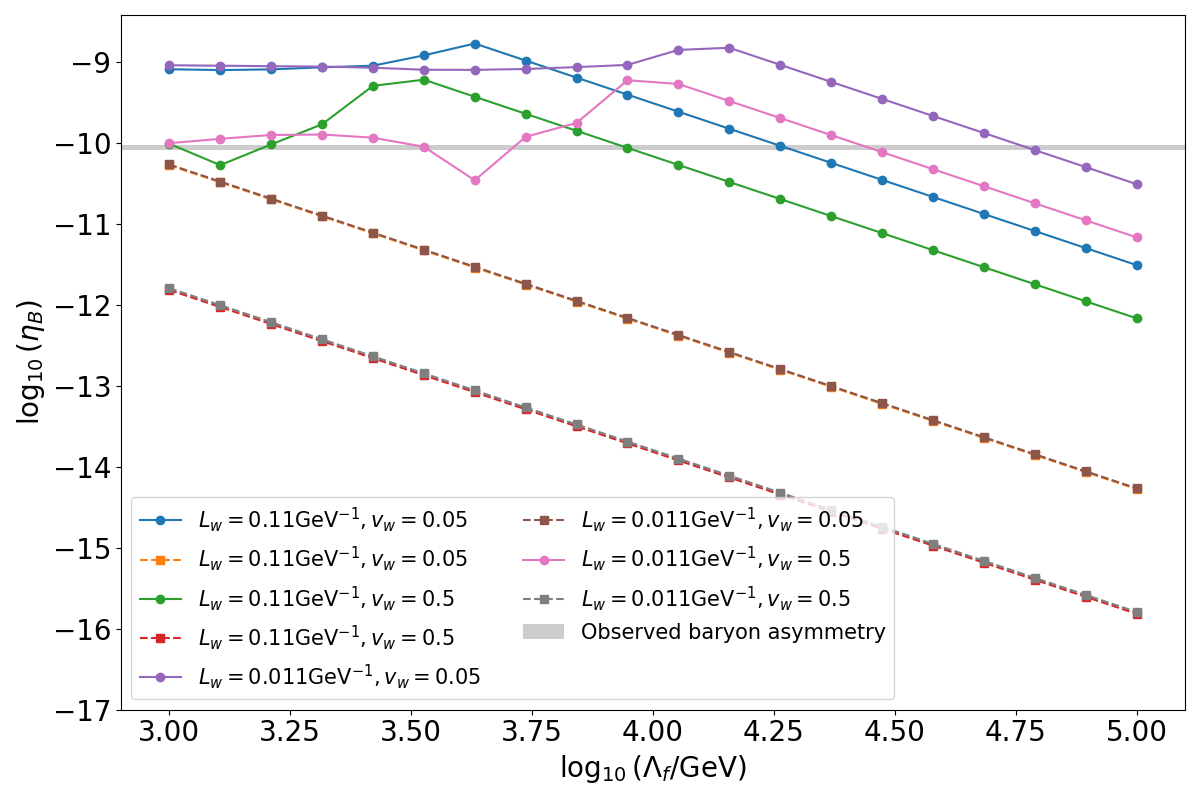}
\caption{Dependence of the baryon asymmetry on the cutoff scale $\Lambda_f$. Solid curves correspond to the baryon asymmetry ($\eta_B=n_B/s$) obtained from numerically solving Eqs.~(\ref{eq:mu5-final},\ref{eq:Bxfinal9},\ref{eq:Byfinal10},\ref{eq:muB-final11}). The final evolutionary outcome is that over a long period of time $t=800~\mathrm{GeV}^{-1}$, all situations reach a saturation point. Dashed curves show the standard EWBG prediction computed from the semi-analytic expression of Eq.~\eqref{YB1811}. The black horizontal line represents the observed value of the baryon asymmetry $(8.2-9.4)\times10^{-11}$~\cite{ParticleDataGroup:2018ovx}. }
\label{eta_CME_EWBG}
\end{figure}

With four representative parameter choices of the wall thickness and velocity, in figure~\ref{eta_CME_EWBG} we present the dependence of the baryon asymmetry on the cutoff scale $\Lambda_f$ in scenarios with and without CME. The dashed curves correspond to the standard EWBG results obtained from the semi-analytic expression of Ref.~\cite{DeVries:2018aul}, where the baryon asymmetry scales as $\eta_B\propto1/\Lambda_f^2$, reflecting the suppression of the CPV source by the high cutoff. 
The solid curves show the result of the full numerical solution of the coupled transport and magnetohydrodynamic equations, evaluated at a late time $t=t_{max}=800\mathrm{GeV}^{-1}$. In this case, the CME induces a dynamical amplification of magnetic fields, which in turn enhances the chiral chemical potential and the baryon asymmetry through anomaly-induced feedback. As a consequence, the suppression with increasing $\Lambda_f$ is significantly weakened, and the baryon asymmetry can be enhanced by several orders of magnitude compared to the standard expectation. The magnitude of this enhancement depends sensitively on the bubble-wall properties. Thinner walls and smaller wall velocities favor stronger chiral asymmetry and more efficient CME-driven amplification.
Figure~\ref{eta_CME_EWBG} thus demonstrates that magnetic-field back-reaction can qualitatively modify the cutoff-scale dependence of EWBG, opening new regions of parameter space compatible with the observed baryon asymmetry.

\begin{figure}[!htp]
\centering
\includegraphics[width=1.0\linewidth]{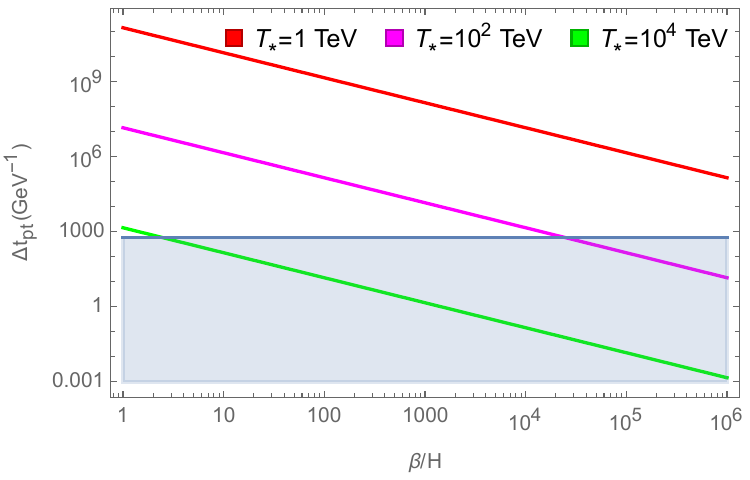}
\caption{Bubble expansion time $\Delta t_{pt}$ as a function of phase transition inverse duration $\beta/H$ for representative temperatures $T_{\star}=1, 10^2, 10^4\;\mathrm{TeV}$. The curves are obtained from $\Delta t_{pt}=[(\beta/H)H(T)]^{-1}$ with the radiation dominated expansion rate $H(T)$. The gray region is the parameter space used to evolve the magnetic field and chemical potential equations.}
\label{deltatpt}
\end{figure}

We finally present the bubble wall evolution time scale in Fig.~\ref{deltatpt}, which shows that for a fast phase transition with $\beta/H_\star \gtrsim\mathcal{O}(10^5)$ at $T_\star= 100$ TeV might not allow enough time for the CME effect, and then our mechanism would not yield a strong enhancement of the baryon number density. For the $T_\star=10^4$ TeV phase transition, there is no time for the CME effect to work.  

\section{Conclusions and Discussions}

Our results highlight the importance of consistently treating anomalous transport and magnetic-field dynamics during a first-order electroweak phase transition. The inclusion of CME-induced MHD evolution provides a viable pathway to reconcile successful baryogenesis with stringent EDM constraints, and it suggests that primordial magnetic fields may play an essential dynamical role in baryon asymmetry generation.

In this work, we show that the variation of the CPV source around the bubble drives a CME-induced growth of helical magnetic fields. The amplified magnetic field configurations, in turn, enhance
the chiral chemical potential of the lepton through
The anomaly-induced contribution, establishing a positive feedback loop that can raise the lepton asymmetry and the magnetic field, yields the baryon asymmetry well beyond the expectation of conventional EWBG scenarios without MHD backreaction.
  
We present the dependence of the BAU on the bubble-wall properties and the CPV scale qualitatively. In particular, smaller wall velocities and thinner walls generally favor a larger net baryon yield because they prolong the effective interaction time and strengthen the wall-localized CP-violating source, while the CME-driven amplification partially compensates the suppression at larger velocities. As a result, the standard scaling of $\eta_B$ with the CP violation scale $\Lambda_f$ is significantly weakened once the MHD evolution is included, opening up regions of parameter space where the observed baryon asymmetry can be reproduced even for $\Lambda_f$ in the range of $\mathcal{O}(10^2)$ TeV.

For the mechanism proposed in this work to be active, the phase transition should occur below $T_\star\lesssim 10$ TeV, and we need to require the inverse duration $\beta/H_\star \lesssim \mathcal{O}(10^5)$ for phase transition temperature $T_\star=100$ TeV. Part of the parameter spaces of $10^2$ TeV $\lesssim T_\star\lesssim 10^4$ TeV indeed have already been excluded by the LIGO observation for strong phase transition~\cite{Romero:2021kby}. Future gravitational wave detections, such as LISA, TianQin, and Taiji, might probe other viable parameter spaces under study. 
We further note that a more comprehensive study requires numerical simulations on the coupled system of the background scalar field, fluid, and magnetic fields.

\section{Acknowledgements}

We are grateful to Kohei Kamada for helpful discussions on the CME effect.
This work is supported by the National Natural Science Foundation of China (NSFC) under Grants Nos.2322505, 12547101. We also acknowledge the Chongqing Natural Science Foundation under Grant No. CSTB2024NSCQJQX0022 and Chongqing Talents: Exceptional Young Talents Project No. cstc2024ycjhbgzxm0020.

\bibliography{./reference.bib}

\clearpage
\appendix
% \onecolumngrid
%\section{Supplementary material}
\section{Appendix A: Calculation of Transport Coefficients}\label{appa}
In this appendix, we provide additional details for calculating transport coefficients~\cite{Xie:2020wzn}.
\begin{multline}
\Gamma_M = \frac{3}{\pi^2 T_n^3} |m_\tau|^2 
\int_0^\infty \frac{k^2\, dk}{\omega_L \omega_R}
\times {\rm Im}\Bigg[
 \frac{h_F(\epsilon_R)+h_F(\epsilon_L)}{\epsilon_R+\epsilon_L}\\
   (\epsilon_L\epsilon_R+k^2) 
 -\frac{h_F(\epsilon_R^*)+h_F(\epsilon_L)}{\epsilon_R^*-\epsilon_L}
   (\epsilon_L\epsilon_R^*-k^2)
\Bigg].
\end{multline}

and
\vspace{-0.1cm}
\begin{align}
\Gamma_Y &= \frac{3y_\tau^2}{4\pi^3T_n^2}
 (m_h^2 - m_{\tau_L}^2 - m_{\tau_R}^2)
 \int_{m_{\tau_R}}^\infty d\omega_R \, h_F(\omega_R) \notag \\
&\quad \times \ln\!\Bigg[
   \frac{e^{\omega_R/T_n}+e^{\omega_-/T_n}}
        {e^{\omega_R/T_n}+e^{\omega_+/T_n}} \,
   \frac{e^{\omega_+/T_n}-1}
        {e^{\omega_-/T_n}-1}
 \Bigg] \notag \\
&\quad + \frac{3\zeta_3}{32\pi^3}\, g^2 y_\tau^2 T_n
   \ln\!\left(\frac{8T_n^2}{m_{\tau_L}^2}\right) .
\label{Gamma_Y}
\end{align}
where $\zeta_3 \approx 1.202$, $m_{\tau_L,\tau_R,h}^2$ are short for the thermal masses:
\begin{align}
&{\rm Re}[\delta m_h^2(T)] = 
\left[\frac{3}{16}g^2+\frac{1}{16}g'^2 \right. \nonumber\\
&\quad \left.
   +\frac{1}{12}\sum_j\left(y_{e^j}^2
   +3y_{u^j}^2+3y_{d^j}^2\right)\right]T^2,\\[6pt]
&{\rm Re}[\delta m_{\tau_L}^2(T)] = 
\left(\frac{3}{32}g^2+\frac{1}{32}g'^2+\frac{1}{16}y_\tau^2\right)T^2,
   \nonumber\\[6pt]
&{\rm Re}[\delta m_{\tau_R}^2(T)] = 
\left(\frac{1}{8}g'^2+\frac{1}{8}y_\tau^2\right)T^2 .
\label{thermalmass}
\end{align}
and where
\begin{multline}
\omega_\pm=\frac{1}{2m_{\tau_R}^2}
\Big[\omega_R\big|m_h^2+m_{\tau_R}^2-m_{\tau_L}^2\big| \\
\pm \sqrt{\,\big(\omega_R^2-m_{\tau_R}^2\big)
\big(m_{\tau_R}^2-(m_{\tau_L}+m_h)^2\big)} \\
\times \sqrt{\big(m_{\tau_R}^2-(m_{\tau_L}-m_h)^2\big)}\;\Big].
\end{multline}

The $\omega_{R/L}$ and $\epsilon_{R/L}$ functions are
\begin{equation}
\begin{aligned}
&\omega_{R/L}(\mathbf{k})=
\sqrt{|\mathbf{k}|^2+{\rm Re}[\delta m_{\tau_{R/L}}^2(T)]},\\
&\epsilon_{R/L}(\mathbf{k})=
\omega_{R/L}(\mathbf{k})-i\Gamma_{\tau_{R/L}} .
\end{aligned}
\end{equation}
where $\Gamma_{\tau_R}\approx0.002T_n$~\cite{Elmfors:1998hh}. The $n_F$ and $h_F$ functions are
\begin{equation}
    \begin{aligned}
       &n_F(k_0)=\frac{1}{e^{k_0/T_n}+1},\\
       &h_F(k_0)=\frac{e^{k_0/T_n}}{(e^{k_0/T_n}+1)^2}.
    \end{aligned}
\end{equation}
The $k_i$ coefficients are given by~\cite{deVries:2017ncy, Fuchs:2020pun}
\begin{equation}
    \begin{aligned}
        &k_i=\tilde{k}_i\frac{c_{B/F}}{\pi^2}\int_{a_i}^{\infty}dx\frac{xe^x}{(e^x\mp1)^2}\sqrt{x^2-a_{i}^2},
    \end{aligned}
\end{equation}
where the upper/lower sign is for bosons/fermions, $c_{B/F}=3$ or 6 for bosons/fermions, $\tilde{k}_i$ is the physical degrees of freedom in a multiplet, in our paper $\tilde{k}_h=4, \tilde{k}_{\tau_L}=2, \tilde{k}_{\tau_R}=1$, and $a_i=\sqrt{Re[\delta m_i^2(T_n)]/T_n}$. 

Finally, the factor $J_{\tau}$ in the CPV source of Eq.~\ref{eq:SCP_general} is given by~\cite{Lee:2004we, Fuchs:2020pun, Cirigliano:2006wh}
\begin{align}
J_\tau &= \int_0^\infty \frac{k^2\,dk}{\omega_L \omega_R}
\,{\rm Im}\Bigg[
   \frac{n_F(\epsilon_L)-n_F(\epsilon_R^*)}{(\epsilon_L-\epsilon_R^*)^2}
   \big(\epsilon_L\epsilon_R^*-k^2\big) \nonumber\\
&\quad 
+ \frac{n_F(\epsilon_L)+n_F(\epsilon_R)}{(\epsilon_L+\epsilon_R)^2}
   \big(\epsilon_L\epsilon_R+k^2\big)
\Bigg].
\label{Jtau}
\end{align}

\section{Appendix B: Coupled evolution of chiral asymmetry and magnetic fields}
\label{sec:mu5-B-eom}

In this Appendix, we derive the set of coupled evolution equations for the chiral
chemical potential $\mu_5$ and the magnetic field $\boldsymbol{B}$ in a conducting
plasma, including the effects of the chiral anomaly and CME.  We work in flat space with metric
signature $(+,-,-,-)$ and adopt natural units $c=\hbar=k_{Boltz}=1$.

The axial current is defined as
\begin{equation}
  J_5^\mu \equiv \bigl( n_5,\,\boldsymbol{J}_5 \bigr),
\end{equation}
where $n_5$ is the chiral number density (right-handed minus $\tau$ left-handed fermions $\tau$).
In a hydrodynamic description and to leading order in gradients, one may write
\begin{equation}
  J_5^\mu
  = n_5 u^\mu - D_{\tau}\,\partial^\mu \mu_5 + \cdots,
  \label{eq:J5-mu}
\end{equation}
where $u^\mu$ is the fluid four-velocity satisfying $u_\mu u^\mu = 1$,
$\mu_5$ is the chiral chemical potential.

The divergence of the axial current receives an anomalous contribution from
the electromagnetic fields, as well as chirality-flipping interactions, and
CPV sources.  We parametrize this as
\begin{equation}
  \partial_\mu J_5^\mu
  = \frac{g'^2}{16\pi^2}\,\boldsymbol{E}\!\cdot\!\boldsymbol{B}
    - \Gamma_{\rm tot}\,\mu_5
    + S_{\rm CP},
  \label{eq:anomaly-master}
\end{equation}

We relate the charge  density and chemical potential in the relativistic approximation via~\cite{Kamada:2016eeb}
\begin{equation}
  n_5
  \simeq \frac{T^2}{6}\,\mu_5,
  \label{eq:n5-mu5}
\end{equation}
Using Eqs.~\eqref{eq:J5-mu} and \eqref{eq:n5-mu5} we obtain
\begin{equation}
  \boldsymbol{J}_5
  = n_5\boldsymbol{v} - D_{\tau}\,\boldsymbol{\nabla}\mu_5
  = \frac{T^2}{6}\,\mu_5\,\boldsymbol{v}
    - D_{\tau}\,\boldsymbol{\nabla}\mu_5,
  \label{eq:J5-3vec}
\end{equation}
where $\boldsymbol{v}$ is the bulk velocity.

In terms of three-vectors, the anomaly equation \eqref{eq:anomaly-master} can then be
written as
\begin{equation}
  \frac{\partial n_5}{\partial t}
  + \boldsymbol{\nabla}\!\cdot\!\boldsymbol{J}_5
  = \frac{g'^2}{16\pi^2}\,\boldsymbol{E}\!\cdot\!\boldsymbol{B}
    - \Gamma_{\rm tot}\,\mu_5
    + S_{\rm CP}.
  \label{eq:anomaly-3vec}
\end{equation}
Using Eq.~\eqref{eq:n5-mu5},
\begin{equation}
  \frac{\partial n_5}{\partial t}
  = \frac{T^2}{6}\,\frac{\partial\mu_5}{\partial t}.
\end{equation}
Moreover,
\begin{equation}
  \boldsymbol{\nabla}\!\cdot\!\boldsymbol{J}_5
  = \boldsymbol{\nabla}\!\cdot\!\biggl(
      \frac{T^2}{6}\,\mu_5\,\boldsymbol{v}
      - D_{\tau}\,\boldsymbol{\nabla}\mu_5
    \biggr).
\end{equation}
Substituting these relations into Eq.~\eqref{eq:anomaly-3vec} yields
\begin{equation}
  \frac{T^2}{6}\,\frac{\partial\mu_5}{\partial t}
  + \boldsymbol{\nabla}\!\cdot\!\biggl(
      \frac{T^2}{6}\,\mu_5\,\boldsymbol{v}
    \biggr)
  - D_{\tau}\,\boldsymbol{\nabla}^2\mu_5
  = \frac{g'^2}{16\pi^2}\,\boldsymbol{E}\!\cdot\!\boldsymbol{B}
    - \Gamma_{\rm tot}\,\mu_5
    + S_{\rm CP}.
  \label{eq:mu5-eq-pre}
\end{equation}

We now specialize in the planar-wall geometry relevant for a first-order
phase transition.  The wall is taken to lie in the $x$--$y$ plane and
is translationally invariant along these directions.  All quantities
therefore depend only on time $t$ and the coordinate $z$ perpendicular
to the wall,
\begin{equation}
  \frac{\partial}{\partial x}
  = \frac{\partial}{\partial y} = 0, \qquad
  f = f(t,z).
\end{equation}
In addition, we assume the plasma is approximately incompressible,
\begin{equation}
  \boldsymbol{\nabla}\!\cdot\!\boldsymbol{v} = 0.
\end{equation}
In one dimension, this implies
\begin{equation}
  \frac{\partial v_z}{\partial z} = 0
  \quad \Rightarrow \quad
  v_z = \text{const.}
\end{equation}
Adopting the bubble-wall rest frame, the plasma flows past the wall
with a velocity $v_z = -v_w$, where $v_w>0$ is the wall velocity in
the plasma frame.

Under these assumptions, the divergence term in Eq.~\eqref{eq:mu5-eq-pre}
reduces to
\begin{equation}
  \boldsymbol{\nabla}\!\cdot\!\biggl(
     \frac{T^2}{6}\,\mu_5\,\boldsymbol{v}
  \biggr)
  =
  \frac{\partial}{\partial z}
  \biggl(
     \frac{T^2}{6}\,\mu_5\,v_z
  \biggr)
  = \frac{T^2}{6}\,v_z\,\frac{\partial\mu_5}{\partial z},
\end{equation}
where we used $\partial_z v_z = 0$.  The Laplacian becomes
\begin{equation}
  \boldsymbol{\nabla}^2\mu_5
  = \frac{\partial^2\mu_5}{\partial z^2}.
\end{equation}
Therefore, Eq.~\eqref{eq:mu5-eq-pre} takes the form
\begin{equation}
  \frac{T^2}{6}\,\frac{\partial\mu_5}{\partial t}
  + \frac{T^2}{6}\,v_z\,\frac{\partial\mu_5}{\partial z}
  - D_{\tau}\,\frac{\partial^2\mu_5}{\partial z^2}
  = \frac{g'^2}{16\pi^2}\,\boldsymbol{E}\!\cdot\!\boldsymbol{B}
    - \Gamma_{\rm tot}\,\mu_5
    + S_{\rm CP}.
\end{equation}
Multiplying by $6/T^2$ for convenience and using $v_z = -v_w$ we obtain
\begin{equation}
  \frac{\partial\mu_5}{\partial t}
  - v_w\,\frac{\partial\mu_5}{\partial z}
  - D_{\rm eff}\,\frac{\partial^2\mu_5}{\partial z^2}
  = \alpha\,\boldsymbol{E}\!\cdot\!\boldsymbol{B}
    - \tilde{\Gamma}_{\rm tot}\,\mu_5
    + \tilde{S}_{\rm CP}\;.
  \label{eq:mu5-final1}
\end{equation}

%-------------------------------------------------------------
% \subsection{Maxwell equations with the chiral magnetic effect}
% \label{subsec:Maxwell-CME}

We now derive the evolution equation for the magnetic field, including the
chiral magnetic effect.  In the MHD regime, where displacement currents
can be neglected, the relevant Maxwell equations are
\begin{align}
  &\frac{\partial\boldsymbol{B}}{\partial t}
  = -\,\boldsymbol{\nabla}\times\boldsymbol{E},
  \label{eq:Faraday}\\[0.5em]
  &\boldsymbol{\nabla}\times\boldsymbol{B}
  = \,\boldsymbol{J},
  \label{eq:Ampere}
\end{align}
supplemented by the constraint $\boldsymbol{\nabla}\!\cdot\!\boldsymbol{B}=0$.
The generalized Ohm law, including the CME, reads
\begin{equation}
  \boldsymbol{J}
  = \sigma\bigl(\boldsymbol{E} + \boldsymbol{v}\times\boldsymbol{B}\bigr)
    + \frac{\alpha_Y}{\pi}\,\mu_5\,\boldsymbol{B},
  \label{eq:Ohm-CME}
\end{equation}
Where $\sigma$ is the electric conductivity of the plasma.

Using Amp\`ere's law \eqref{eq:Ampere}, 
we can solve Eq.~\eqref{eq:Ohm-CME} for the electric field
\begin{equation}
  \boldsymbol{E}
  = \frac{1}{\sigma}\,\boldsymbol{\nabla}\times\boldsymbol{B}
    - \boldsymbol{v}\times\boldsymbol{B}
    - \frac{\alpha_Y}{\pi\sigma}\,\mu_5\,\boldsymbol{B}.
  \label{eq:E-from-Ohm}
\end{equation}
Substituting Eq.~\eqref{eq:E-from-Ohm} into Faraday's law \eqref{eq:Faraday}
we obtain
\begin{align}
  \frac{\partial\boldsymbol{B}}{\partial t}
  &= -\,\boldsymbol{\nabla}\times\boldsymbol{E}
   = -\,\boldsymbol{\nabla}\times
      \biggl[
        \frac{1}{\sigma}\,\boldsymbol{\nabla}\times\boldsymbol{B}
        - \boldsymbol{v}\times\boldsymbol{B}
        - \frac{\alpha_Y}{\pi\sigma}\,\mu_5\,\boldsymbol{B}
      \biggr]
  \nonumber\\[0.5em]
  &= -\frac{1}{\sigma}\,\boldsymbol{\nabla}\times
       (\boldsymbol{\nabla}\times\boldsymbol{B})
     + \boldsymbol{\nabla}\times(\boldsymbol{v}\times\boldsymbol{B})
     + \frac{\alpha_Y}{\pi\sigma}\,
       \boldsymbol{\nabla}\times(\mu_5 \boldsymbol{B}) .
\end{align}
Using the vector identity
\begin{equation}
  \boldsymbol{\nabla}\times(\boldsymbol{\nabla}\times\boldsymbol{B})
  = \boldsymbol{\nabla}(\boldsymbol{\nabla}\!\cdot\!\boldsymbol{B})
    - \boldsymbol{\nabla}^2\boldsymbol{B}
  = -\,\boldsymbol{\nabla}^2\boldsymbol{B},
\end{equation}
where we used $\boldsymbol{\nabla}\!\cdot\!\boldsymbol{B}=0$,
We finally arrive at the MHD equation
\begin{equation}
  \frac{\partial\boldsymbol{B}}{\partial t}
  = \frac{1}{\sigma}\,\boldsymbol{\nabla}^2\boldsymbol{B}
    + \boldsymbol{\nabla}\times(\boldsymbol{v}\times\boldsymbol{B})
    + \frac{\alpha_Y}{\pi\sigma}\,
      \boldsymbol{\nabla}\times(\mu_5 \boldsymbol{B}) .
  \label{eq:B-vector-eq}
\end{equation}
The first term on the right-hand side describes the resistive diffusion of the
magnetic field, the second term encodes plasma advection and stretching of
field lines, and the third term is the CME-induced contribution, which can
Lead to an instability and exponential growth of helical magnetic fields.
We now rewrite Eq.~\eqref{eq:B-vector-eq} in component form for the same
one-dimensional planar geometry.  We
take
\begin{equation}
\begin{aligned}
  &\boldsymbol{B} = \bigl( B_x(t,z),\,B_y(t,z),\,B_z(t,z) \bigr),\\
  &\boldsymbol{v} = \bigl( v_x(z),\,v_y(z),\,v_z(z) \bigr),
\end{aligned}
\end{equation}
with $\partial_x=\partial_y=0$.

The Laplacian reduces to
\begin{equation}
  \boldsymbol{\nabla}^2\boldsymbol{B}
  = \frac{\partial^2\boldsymbol{B}}{\partial z^2}
  = \biggl(
      \frac{\partial^2 B_x}{\partial z^2},
      \frac{\partial^2 B_y}{\partial z^2},
      \frac{\partial^2 B_z}{\partial z^2}
    \biggr).
\end{equation}

For the advection term, we first compute
\begin{equation}
  \boldsymbol{v}\times\boldsymbol{B}
  = \begin{pmatrix}
      v_y B_z - v_z B_y \\
      v_z B_x - v_x B_z \\
      v_x B_y - v_y B_x
    \end{pmatrix},
\end{equation}
and then its curl:
\begin{equation}
  \boldsymbol{\nabla}\times(\boldsymbol{v}\times\boldsymbol{B})
  = \begin{pmatrix}
      \partial_y (v_x B_y - v_y B_x)
        - \partial_z (v_z B_x - v_x B_z) \\[0.3em]
      \partial_z (v_y B_z - v_z B_y)
        - \partial_x (v_x B_y - v_y B_x) \\[0.3em]
      \partial_x (v_z B_x - v_x B_z)
        - \partial_y (v_y B_z - v_z B_y)
    \end{pmatrix}.
\end{equation}
Using $\partial_x=\partial_y=0$ this simplifies to
\begin{align}
  \bigl[\boldsymbol{\nabla}\times(\boldsymbol{v}\times\boldsymbol{B})\bigr]_x
  &= -\,\frac{\partial}{\partial z}\bigl( v_z B_x - v_x B_z \bigr),\\[0.3em]
  \bigl[\boldsymbol{\nabla}\times(\boldsymbol{v}\times\boldsymbol{B})\bigr]_y
  &= \frac{\partial}{\partial z}\bigl( v_y B_z - v_z B_y \bigr),\\[0.3em]
  \bigl[\boldsymbol{\nabla}\times(\boldsymbol{v}\times\boldsymbol{B})\bigr]_z
  &= 0.
\end{align}

Similarly, for the CME term, we compute
\begin{equation}
  \mu_5 \boldsymbol{B}
  = \bigl( \mu_5 B_x,\;\mu_5 B_y,\;\mu_5 B_z \bigr),
\end{equation}
and hence
\begin{equation}
  \boldsymbol{\nabla}\times(\mu_5 \boldsymbol{B})
  =
  \begin{pmatrix}
     -\partial_z(\mu_5 B_y) \\[0.3em]
      \partial_z(\mu_5 B_x) \\[0.3em]
      0
  \end{pmatrix}.
\end{equation}

The $x$- and $y$-components of Eq.~\eqref{eq:B-vector-eq} therefore read
\begin{align}
  \frac{\partial B_x}{\partial t}
  &= \frac{1}{\sigma}\,\frac{\partial^2 B_x}{\partial z^2}
     - \frac{\partial}{\partial z}\bigl( v_z B_x - v_x B_z \bigr)
     - \frac{\alpha_Y}{\pi\sigma}\,
       \frac{\partial}{\partial z}(\mu_5 B_y),
  \label{eq:Bx-general}\\[0.5em]
  \frac{\partial B_y}{\partial t}
  &= \frac{1}{\sigma}\,\frac{\partial^2 B_y}{\partial z^2}
     + \frac{\partial}{\partial z}\bigl( v_y B_z - v_z B_y \bigr)
     + \frac{\alpha_Y}{\pi\sigma}\,
       \frac{\partial}{\partial z}(\mu_5 B_x).
  \label{eq:By-general}
\end{align}
The $z$-component reduces to purely diffusive evolution,
\begin{equation}
  \frac{\partial B_z}{\partial t}
  = \frac{1}{\sigma}\,\frac{\partial^2 B_z}{\partial z^2}.
\end{equation}
On the other hand, the constraint
$\boldsymbol{\nabla}\!\cdot\!\boldsymbol{B}=0$ implies
$\partial_z B_z=0$ in the one-dimensional setup, so that $B_z$ is independent
of $z$.  If we further assume $B_z$ does not evolve in time, it may be treated
as a constant background and dropped from the diffusion equation.

Using the product rule, the advection terms in Eqs.~\eqref{eq:Bx-general}
and \eqref{eq:By-general} can be rewritten as
\begin{align}
  \frac{\partial}{\partial z}\bigl( v_z B_x - v_x B_z \bigr)
  &= v_z\,\frac{\partial B_x}{\partial z}
     - B_z\,\frac{\partial v_x}{\partial z},\\[0.3em]
  \frac{\partial}{\partial z}\bigl( v_y B_z - v_z B_y \bigr)
  &= B_z\,\frac{\partial v_y}{\partial z}
     - v_z\,\frac{\partial B_y}{\partial z}.
\end{align}
Inserting these into Eqs.~\eqref{eq:Bx-general} and \eqref{eq:By-general}
we obtain
\begin{align}
  \frac{\partial B_x}{\partial t}
  &= \frac{1}{\sigma}\,\frac{\partial^2 B_x}{\partial z^2}
     - v_z\,\frac{\partial B_x}{\partial z}
     + B_z\,\frac{\partial v_x}{\partial z}
     - \frac{\alpha_Y}{\pi\sigma}\,
       \frac{\partial}{\partial z}(\mu_5 B_y),
  \label{eq:Bx-incomp}\\[0.5em]
  \frac{\partial B_y}{\partial t}
  &= \frac{1}{\sigma}\,\frac{\partial^2 B_y}{\partial z^2}
     - v_z\,\frac{\partial B_y}{\partial z}
     + B_z\,\frac{\partial v_y}{\partial z}
     + \frac{\alpha_Y}{\pi\sigma}\,
       \frac{\partial}{\partial z}(\mu_5 B_x).
  \label{eq:By-incomp}
\end{align}
In the simplest case where the transverse velocities vanish,
$v_x = v_y = 0$, and $v_z = -v_w$ is a constant, the gradient terms
$\partial_z v_x$ and $\partial_z v_y$ disappear and Eqs.~\eqref{eq:Bx-incomp}
and \eqref{eq:By-incomp} reduce to
\begin{align}
  \frac{\partial B_x}{\partial t}
  &= \frac{1}{\sigma}\,\frac{\partial^2 B_x}{\partial z^2}
     + v_w\,\frac{\partial B_x}{\partial z}
     - \frac{\alpha_Y}{\pi\sigma}\,
       \frac{\partial}{\partial z}(\mu_5 B_y),
  \label{eq:Bx-final}\\[0.5em]
  \frac{\partial B_y}{\partial t}
  &= \frac{1}{\sigma}\,\frac{\partial^2 B_y}{\partial z^2}
     + v_w\,\frac{\partial B_y}{\partial z}
     + \frac{\alpha_Y}{\pi\sigma}\,
       \frac{\partial}{\partial z}(\mu_5 B_x).
  \label{eq:By-final}
\end{align}
Equations~\eqref{eq:mu5-final}, \eqref{eq:Bx-final}, and \eqref{eq:By-final}
constitute the closed system for the coupled evolution of the chiral chemical
potential and the transverse magnetic field in the planar-wall geometry,
which is used in our numerical analysis.

\section{Appendix C: The baryon asymmetry generation}

Electroweak sphalerons change baryon and lepton numbers with
$\Delta B=\Delta L=n_f$ per topological transition ($n_f=3$).
Near equilibrium the baryon number density $n_B$ obeys (detailed balance)~\cite{Joyce:1994bk, Riotto:1999yt}:
\begin{equation}
\partial n_B \;=\; -\,n_f\,\Gamma_{ws}\,\bigl(\mu_{ws}+\mu^{0}_{ws}\bigr),
\label{eq:db}
\end{equation}
where $\mu_{ws}$ is the chemical potential combination coupled to the
anomalous SU(2) process,
\begin{align}
\mu_{ws}
&= \sum_{i=1}^{n_f}\!\bigl(3\mu_{q_{L,i}}+\mu_{\ell_{L,i}}\bigr)
= \sum_i \!\left(\frac{3n_{q_{L,i}}}{k_{q_L}}+\frac{n_{\ell_{L,i}}}{k_{L}}\right)
\nonumber\\
&\simeq \frac{1}{2}\sum_i\!\bigl(n_{q_{L,i}}+n_{\ell_{L,i}}\bigr)
\equiv \frac{1}{2}\,n_L.
\label{eq:muws-def}
\end{align}
We use the linear response $n_a=k_a\,\mu_a$ with (rescaled) susceptibilities $k_a$.
In the one-step treatment, we take $\mu^{0}_{ws}=0$.

The quark contribution is related to $n_B$ by
\begin{equation}
n_B \;=\; n_B^{L}+n_B^{R} \;=\; 2\,n_B^{L}
\;=\; \frac{2}{3}\sum_i n_{q_{L,i}}
\label{eq:nb-quark}
\end{equation}
(three colors per family).
Using $B-L$ conservation to relate the lepton piece, one obtains the
closure
\begin{equation}
\mu_{ws}\;=\;\frac{1}{2}\sum_i(n_{q_{L,i}}+n_{\ell_{L,i}})
\;=\; \frac{\mathcal{R}}{3}\,n_B,
\label{eq:R-closure}
\end{equation}
In the wall frame, we define
$\partial n_B\equiv \partial_t n_B-v_w\,n_B' - D_q\,n_B''$ and obtain
\begin{equation}
\partial_t n_B-v_w n_B' - D_q n_B'' \;=\; -\,\Gamma_{ws}\!\left(\frac{3}{2}\,n_L^{0} + \mathcal{R}\,n_B\right).
\label{eq:nb-eq}
\end{equation}
Where the constant $\mathcal{R}=15/4$~\cite{DeVries:2018aul} is universal in this framework and generates the familiar
washout coefficient $-15/4$ once we change variables from $n_B$ to $\mu_B$.

To linear order, one may write $n_B\approx T^2\mu_B/6$.
Then, Eq.~\eqref{eq:nb-eq} yields
\begin{equation}
\partial_t \mu_B - v_w\,\partial_z \mu_B
= D_B\,\partial_z^2\mu_B
-\frac{15}{4} \Gamma_{ws}\,\mu_B.
\label{eq:muB-washout}
\end{equation}
where $D_B=6D_q/T^2$. To obtain the $\mu_5$ source, we must compute the response of $\mu_{ws}$ to a
leptonic chiral bias under the same closure used 
top Yukawa fast, strong sphalerons fast, bottom/charm Yukawas neglected,
$\tau$ sector source~\cite{DeVries:2018aul}. We work on a species basis
\begin{equation}
\mu = (Q, U, T, D, B, L, R, H),
\end{equation}
where $Q$ is the left--handed quark doublet,
$U$ ($D$) are the two light up (down) singlets $(u,c)$ [$(d,s)$],
$T$ and $B$ denote the right-handed top and bottom,
$L$ and $R$ the $\tau$ doublet and singlet, and $H$ the Higgs.
We use zero mass susceptibilities, including color/isospin/family factors:
\begin{eqnarray}
&k_Q=18,\quad k_U=6,\quad k_T=3,\quad
k_D=6,\nonumber\\
\label{eq:ks}
 &    k_{B}=3,\quad
k_L=2,\quad k_R=1,\quad k_H=4\;.
\end{eqnarray}
Fast Yukawa interactions:
\begin{equation}
T=Q+H,\qquad R=L+H,
\label{eq:yukawa-fast}
\end{equation}
so that $\mu_5\equiv R-L=H$ in the $\tau$ sector.
Strong sphaleron equilibrium (three families combined):
\begin{equation}
6Q-2U-T-2D-B=0.
\label{eq:ss}
\end{equation}
Hypercharge neutrality (with the weights implicit in \eqref{eq:ks}):
\begin{equation}
3Q+4U+2T-2D-B-L-R+2H=0.
\label{eq:Y0}
\end{equation}

We minimize the quadratic free energy
\begin{equation}
\begin{aligned}
&\mathcal F(\mu)=\sum_a k_a\,\mu_a^2\\
&=18Q^2+6U^2+3T^2+6D^2+3B^2+2L^2+R^2+4H^2\;,   
\end{aligned}
\end{equation}
subject to the linear constraints \eqref{eq:yukawa-fast}--\eqref{eq:Y0},
treating $H$ as an external parameter.
Eliminating $T$ and $R$ via \eqref{eq:yukawa-fast}, imposing
\eqref{eq:ss} and \eqref{eq:Y0} with Lagrange multipliers, one solves the
resulting linear system and finds the unique minimum
\begin{eqnarray}
&&Q=-\frac{26}{267}H,\quad
U=-\frac{173}{267}H,\quad
D=-\frac{17}{267}H,\\
\label{eq:solution}
   && B=-\frac{17}{267}H,\quad
L=\frac{5}{89}H.
\end{eqnarray}
The weak sphaleron chemical potential combination is
\begin{equation}
\mu_{ws} \;=\; 9Q + L
\;=-\frac{73}{89}\,H
\;= -\frac{73}{89}\,\mu_5\;,
\label{eq:muws-a5}
\end{equation}
which yields
\begin{equation}
a_5\equiv\frac{\partial \mu_{ws}}{\partial \mu_5}= -\frac{73}{89}\;.
\label{eq:a5}
\end{equation}

Combining Eq.~\eqref{eq:R-closure} and Eq.~\eqref{eq:muws-a5}, we have
\begin{equation}
\mu_{ws} \;=\; \frac{R}{3}\,n_B
\;+\; a_5\,\mu_5\;.
\end{equation}
Inserting into Eq.~\eqref{eq:db}, we get the equation of the $\mu_B$:
\begin{align}
\partial_t \mu_B - v_w\,\partial_z \mu_B
=& D_B\,\partial_z^2\mu_B
-\frac{15}{4}\Gamma_{ws}\,\mu_B\nonumber\\
&+ C_B\,\Gamma_{ws}\,\mu_5\;,
\label{eq:muB-final}
\end{align}
where $
C_B = -n_f\,a_5
\;=\; \frac{219}{89}$.
The baryon asymmetry can be numerically obtained as:
\begin{equation}
    \eta_B=\frac{n_B}{s}=\frac{T^2\mu_B}{6s}.
    \label{BAU}
\end{equation}

\end{document}